\documentclass{ceurart}
\usepackage[utf8]{inputenc}
\usepackage{todonotes}
\usepackage[all]{nowidow}

\title{Online Information Retrieval Evaluation using the STELLA Framework}

\author[1]{Timo Breuer}[orcid=0000-0002-1765-2449,email=timo.breuer@th-koeln.de]
\author[2]{Narges Tavakolpoursaleh}[orcid=0000-0001-9324-3252,email=narges.tavakolpoursaleh@gesis.org]
\author[3]{Johann Schaible}[orcid=0000-0002-5441-7640,email=j.schaible@eufh.de]
\author[2]{Daniel Hienert}[orcid=0000-0002-2388-4609,email=daniel.hienert@gesis.org]
\author[1]{Philipp Schaer}[orcid=0000-0002-8817-4632,email=philipp.schaer@th-koeln.de,url=https://ir.web.th-koeln.de]
\author[4]{Leyla Jael Castro}[orcid=0000-0003-3986-0510,email=ljgarcia@zbmed.de]

\address[1]{TH Köln -- University of Applied Sciences,
  Cologne, Germany}

\address[2]{GESIS -- Leibniz Institute for the Social Sciences, Cologne, Germany}  

\address[3]{EU|FH -- University of Applied Sciences, Bruehl, Germany}

\address[4]{ZB~MED -- Information Centre for Life Sciences, Cologne, Germany}

\copyrightyear{2021}
\copyrightclause{Copyright for this paper by its authors.
  Use permitted under Creative Commons License Attribution 4.0
  International (CC BY 4.0).}

\conference{Preprint for Information Retrieval Meeting 2022}

\begin{document}

\maketitle





\section{Abstract}
\textbf{Introduction.}
Involving users in early phases of software development has become a common strategy as it enables developers to consider user needs from the beginning. Once a system is in production, new opportunities to observe, evaluate and learn from users emerge as more information becomes available. Gathering information from users to continuously evaluate their behavior is a common practice for commercial software, while the Cranfield paradigm remains the preferred option for Information Retrieval (IR) and recommendation systems in the academic world. Here we introduce the Infrastructures for Living Labs STELLA project which aims to create an evaluation infrastructure allowing experimental systems to run along production web-based academic search systems with real users. STELLA combines user interactions and log files analyses to enable  large-scale A/B experiments for academic search.

\textbf{Methods.}
The STELLA evaluation infrastructure provides an online reproducible environment allowing developers and researchers to work together to produce and evaluate new retrieval and recommendation approaches for existing IR. STELLA integrates itself to a production system, allows experimental systems to run along the production one, and evaluates the performance of those experimental systems using real-time information coming from the regular users of the system. The production system acts as a baseline that experimental systems try to outperform. Our experimental setup uses interleaving, i.e. it combines experimental results with those from the corresponding baseline systems. STELLA gathers information on user interactions and provides statistics useful to developers and researchers. STELLA architecture comprises three main elements: (i) micro-services corresponding to experimental systems, (ii) a multi-container application (MCA) bundling together all the participant experimental systems, and (iii) a central server to manage participant and production systems, and to provide feedback.

\textbf{Results.}
STELLA was the technological and methodological foundation of the CLEF 2021 Living Labs for Academic Search (LiLAS) lab. LiLAS aimed to strengthen the concept of user-centric living labs for academic search with two separated evaluation rounds of 4 weeks each. LiLAS integrated STELLA into two academic search systems: LIVIVO (for the task of ranking documents wrt a head query) and GESIS Search (for the task of ranking datasets wrt a reference document). We evaluated nine experimental systems contributed by three participating groups. Overall, we consider our lab as a successful advancement to previous living lab experiments. We were able to exemplify the benefits of fully dockerized systems delivering results for arbitrary results on-the-fly. Furthermore, we could confirm several previous findings, for instance the power laws underlying the click distributions.

\section{Introduction}

Involving users in the early phases of software development -- in the form of user experience analysis and prototype testing -- has become common whenever some degree of user interaction is required. This allows developers to consider the users' needs from the beginning, making it easier for users to adopt a new system or adapt to a new version. However, once a system is put in place, new opportunities to observe, evaluate and learn from users become possible as more information becomes available. For instance, it becomes possible to observe and record interaction patterns and answer questions like (a) how long does it take a user to find a particular button, (b) how does the user reacts to different options or (c) what are the most common paths used to achieve a goal, whether this goal is buying clothes or finding an article or dataset relevant for research. All this information can be tracked, stored, analyzed, and evaluated, making systems more attractive and easier for users. Gathering information from users in order to continuously evaluate their interaction and learn more from their needs is a common practice for commercial software as knowing their users and their behavior allows them to predict their actions, offer them better products that they are likely to buy and, therefore, make more profit. Despite the benefits of user-based evaluation, this approach is not yet fully exploited by Information Retrieval systems within the academic world.

Information Retrieval (IR) systems are commonly used in academics as they aim at presenting the most relevant resources from a corpus for an information need. Typical tasks include ranking a series of documents regarding a query or offering recommendations regarding a document already selected as relevant (systems taking care of the latter are called recommendation systems). Traditionally, IR and recommendation systems are used in academics to recover scholarly articles from specialized repositories designed to this end. 
Although IR is an active research area, evaluation remains a challenge in the academic context. One reason for this is that IR evaluation mainly relies on the Cranfield paradigm. In this paradigm, search systems are compared by processing a set of queries or topics based on a standard corpus of documents while trying to produce the best possible results. Results are then evaluated with the help of relevance assessments produced by domain experts. This research method has been established and proven for more than 25 years in international evaluation campaigns such as the Text Retrieval Conference TREC \footnote{\url{https://trec.nist.gov/}} or the Conference and Labs Evaluation Forum CLEF \footnote{\url{https://www.clef-initiative.eu/}}. However, this so-called offline evaluation or shared task principle faces the criticism of drifting away from the actual user needs, as the experts carrying out the evaluations do not correspond to the average users of a platform or correspond only to a particular subgroup.

The Cranfield paradigm has proved to be a good option for IR academic challenges as it reduces the variables that users would introduce. However, academic IR systems face a challenge as more research outcomes are needed nowadays for researchers to have a clear picture. It is no longer enough to retrieve a scientific article without having access to the data it relies on and the software and workflows used to collect and transform such data. It is time for academic IR systems to turn towards more interactive evaluation paradigms, such as those used in industry. Thanks to the online nature of IR systems, system providers and business owners have access to many user interactions on their systems. The interactions can be recorded and evaluated by subsequent log file analysis making large-scale A/B experiments possible. For example, variants of ranking or recommendation algorithms can be offered to users without being aware of it. Furthermore, systems can adapt to multiple user types depending on their observed behavior in the past \cite{DBLP:journals/dbsk/SchaibleBTMWS20}.

The Infrastructures for Living Labs STELLA project \footnote{\url{https://stella-project.org/}} aims to create an evaluation infrastructure allowing retrieval and recommendation systems to run within production web-based academic search systems with real users. STELLA provides an integrated e-research environment, a so-called living lab. The experimental setup differs significantly from classical offline TREC studies. It allows academic researchers to use an evaluation method previously reserved for industrial research or operators of large online platforms. German research infrastructures have sporadically or not at all used such evaluation methods to improve their search systems. However, the potential of living labs and real-world evaluation techniques has been shown in previous CLEF labs such as NewsREEL \footnote{\url{https://www.newsreelchallenge.org/}} and Living Labs for Information Retrieval Evaluation 2015 (LL4IR) or TREC OpenSearch 2016. In the rest of this manuscript, we will introduce the methodology used by the STELLA infrastructure, present some results, and summarize our future work.

\section{Methods}

The STELLA evaluation infrastructure provides an online reproducible environment allowing developers and researchers to work together in order to produce and evaluate new retrieval and recommendation approaches for IR systems already in production. STELLA integrates itself to a production system, allows experimental systems to run along the production one, and evaluates the performance of those experimental systems using real-time information coming from the regular users of the system. The production system acts as a baseline that experimental systems try to outperform. STELLA gathers information on how users react to data coming from both production and experimental systems and provides statistics useful to developers and researchers. One one side, STELLA can be used for online evaluations on academic challenges such as TREC while, on the other side, it also supports IR development teams to decide whether an experimental system should be developed further and later be integrated to the production system to, for instance, improve performance or services \cite{breuer2021}.

\begin{figure}[t]
    \centering
    \includegraphics[width=1\linewidth]{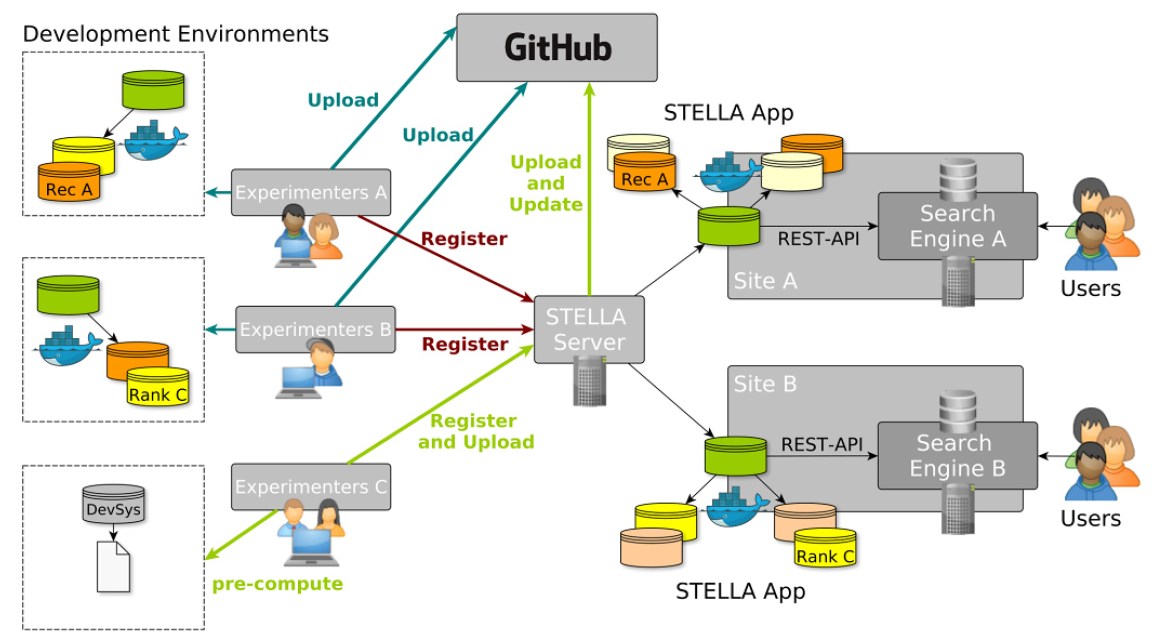}
    \caption{Infrastructure design illustrating how experimental systems can be contributed as Docker-based micro-services to online experiments with real users.}
    \label{fig:stella_architecture}
\end{figure}

Reproducibility is one of the main principles behind the STELLA infrastructure as it is central for any sort of evaluation, whether offline relying on the Cranfield paradigm, or online relying, for instance, on A/B testing. A reproducible environment allows us to validate results obtained by different retrieval approaches and with different experimental setups. In order to enable reproducibility of online experiments on IR systems, STELLA requires full access to the production system and corresponding data supporting the IR capabilities to be evaluated. To avoid misuses of the production system opening up to experimental ones, STELLA requires any experimental system to be containerized, i.e., using Docker\footnote{\url{https://www.docker.com/}}, so that they implement a web service with predefined endpoints, namely ranking or recommendation. STELLA comprises three main elements as seen in the overview of the STELLA architecture provided in Figure \ref{fig:stella_architecture}: (i) micro-services corresponding to experimental systems, (ii) a multi-container application (MCA) bundling together all the participant experimental systems, and (iii) a central server to manage participant and production systems, and to provide feedback. In the following, we describe in more detail the STELLA elements supporting online evaluation with real user information. In addition to the online evaluation, STELLA also supports, mainly for compatibility purposes, offline evaluation in the form of pre-computed systems. As the main focus of this manuscript is the online evaluation, we will not discuss the pre-computed option further. 

\textbf{Micro-services.} Participant experimental systems will deliver information regarding a head query or a recommendation by implementing a REST-based web service with two main endpoints, one for ranking and one for recommendations. To make it easier for participants, we provide templates in our GitHub space. The participants should have access to a detailed description of the corpus behind the production system so they can implement an alternative index that will serve the retrieval requests. The participant systems must be hosted in a Git public repository so they can be integrated to the MCA. The Git repository must be registered at the central dashboard\footnote{\url{https://github.com/stella-project/stella-micro-template}}. 

\textbf{Multi-container application.} The MCA will first run some unit tests on the registered participants. For those participants passing the tests, the MCA will bundle them together and thus act as a single entry point to the infrastructure. One a production system gets a query, it passes it to the MCA which will interleave, i.e., combine, results coming from one of the experimental systems with those from the production one. Using a Team-Draft-Interleaving approach rather than a pure A/B testing prevents bad results far apart from what users are already familiar with. Interleave results also allow the inference of statistically significant results with less data compared to A/B testing. 

\textbf{Central server.} The central server serves four main functionalities: (i) registration of participants, productions systems and administrators, (ii) dashboard service providing visual analytics regarding performance of experimental systems, (iii) storage of feedback data collected from user interactions, and (iv) automated update of the MCA in case new experimental systems are registered.

\section{Results - STELLA at LiLAS 2021}

STELLA was the technological and methodological foundation of the CLEF 2021 Living Labs for Academic Search (LiLAS) lab \cite{schaer2021overview}. 
The LiLAS lab aimed to strengthen the concept of user-centric living labs for academic search. The methodological gap between real-world and lab-based evaluation should be bridged by allowing lab participants to evaluate their retrieval approaches in two real-world academic search systems from life sciences and social sciences. For this purpose LiLAS established two separated evaluation rounds of 4 weeks each and by integrating STELLA into the two academic search systems LIVIVO and GESIS Search. In these two systems two typical academic search tasks were set up to be evaluated within the LiLAS lab, which are ad-hoc retrieval and dataset recommendation. Each participating groups received a set of feedback data after each round; the feedback was also made publicly available on the lab website\footnote{\url{https://th-koeln.sciebo.de/s/OBm0NLEwz1RYl9N}}. Before each round a training phase was offered to allow the participants to build or adapt their systems to the new datasets or click feedback data. As the lab was based on STELLA, participants were allowed to submit results corresponding to their experimental systems in the form of pre-computed runs and Docker containers that were integrated into production systems and generate experimental results in real-time. Both submission types were interleaved with the results provided by the productive systems allowing for a seamless presentation and evaluation. 

The LiLAS lab with the help of STELLA successfully introduced the living lab paradigm with a focus on tasks in the domain of academic search. In total, we evaluated nine experimental systems out of which seven were contributed by three participating groups. In sum, two groups contributed experiments that cover pre-computed rankings and fully dockerized systems at LIVIVO and pre-computed recommendations at GESIS. The GESIS research team contributed another completely dockerized recommendation system. Our experimental setup were based on interleaving experiments that combine experimental results with those from the corresponding baseline systems at LIVIVO and GESIS. In accordance with the living lab paradigm, our evaluations were based on user interactions, i.e. in the form of click feedback.

In addition to established outcome measures of interleaving experiments (Win, Loss, Tie, Outcome), we also account for the meaning of clicks on different SERP elements. In this context, we implement the Reward measure that is the weighted sum of clicks on different elements corresponding to a specific result. Even though most of the experimental systems could not outperform the baseline systems in terms of the overall scores, we see some clear differences between the system performance, which allow us to assess a system's merits more thoroughly, when the evaluations are based on different SERP elements.

Overall, we consider our lab as a successful advancement to previous living lab experiments. We were able to exemplify the benefits of fully dockerized systems delivering results for arbitrary results on-the-fly. Furthermore, we could confirm several previous findings, for instance the power laws underlying the click distributions. Additionally, we were able to conduct more diverse comparison by differentiating between clicks on different SERP elements and accounting for their meaning.

\section{Conclusions}

A key component of the STELLA infrastructure is the integration of experimental ranking and recommendation systems as micro-services that are implemented with the help of Docker into real-world productive retrieval systems. The LiLAS lab was the first test-bed to use this evaluation service and it exemplified some of the benefits resulting from the new infrastructure design. The lab's results included a clear indication that completely dockerized systems can overcome the restrictions of results limited to filtered lists of top-k queries or target items. Significantly more data and click interactions could be logged if the experimental systems delivered results on-the-fly for arbitrary requests of rankings and recommendations. As a consequence, this allowed much more data aggregation in a shorter period of time and provided a solid basis for statistical significance tests.

Another positive side effect of relying on Docker was that the deployment effort for site providers and organizers is considerably reduced. Once the systems are properly described with the corresponding Dockerfile, they can be rebuild on purpose, exactly as the participants and developers intended them to be. Likewise, the entire infrastructure service can be migrated with minimal costs due to Docker. 
If the systems are properly adapted to the required interface and the source code is available in a public repository, the (IR) research community can rely on these artifacts that make the experiments transparent and reproducible.

In the future, it might be helpful to provide participants with open and more transparent baseline systems they can build upon. Some of the pre-computed experimental ranking and recommendations seem to deliver promising results; however, the evaluations need to be interpreted with care due to the sparsity of the available click data. As a next step we focus on the implementation of continuous evaluation principles freed from the time limits of rounds, in order to re-frame the introduced living lab service as an ongoing evaluation challenge.

\section{Acknowledgements}

This work was supported by DFG (project no. 407518790).

\bibliography{references}

\end{document}